\documentclass[%
superscriptaddress,
preprint,
 amsmath,amssymb,
 aps,
]{revtex4-1}

\usepackage{graphicx}% Include figure files
\usepackage{dcolumn}% Align table columns on decimal point
\usepackage{bm}% bold math
\usepackage{xcolor}
\usepackage{hyperref}% add hypertext capabilities
\usepackage{algorithm}
\usepackage{algorithmicx}
\usepackage{algpseudocode}
\usepackage{qcircuit}
\usepackage{braket}
\usepackage{float}
\usepackage{subcaption}
\usepackage{multirow}
\usepackage[frozencache=true,cachedir=.]{minted} 

\setminted[python]{
frame=lines,
framesep=1.5mm,
fontsize=\footnotesize,
linenos,
style=pastie
}

\renewcommand{\figureautorefname}{Figure~\negthinspace}

\renewcommand{\tableautorefname}{Table~\negthinspace}

\newcommand\norm[1]{\left\lVert#1\right\rVert}

\algnewcommand{\algorithmicand}{\textbf{ and }}
\algnewcommand{\algorithmicor}{\textbf{ or }}
\algnewcommand{\OR}{\algorithmicor}
\algnewcommand{\AND}{\algorithmicand}

\begin{document}

\preprint{BNL-UMD}

\title{Quantum Architecture Search via Deep Reinforcement Learning}% Force line breaks with \\

\author{En-Jui Kuo}%
 \email{kuoenjui@umd.edu}
\affiliation{Department of Physics, University of Maryland, College Park, MD 20742, USA}
\affiliation{Joint Quantum Institute, NIST/University of Maryland, College Park, MD 20742, USA}%

\author{Yao-Lung L.\ Fang}%
 \email{leofang@bnl.gov}
\affiliation{Computational Science Initiative, Brookhaven National Laboratory, Upton, NY 11973, USA}%

\author{Samuel Yen-Chi Chen}
\email{ychen@bnl.gov}
\affiliation{Computational Science Initiative, Brookhaven National Laboratory, Upton, NY 11973, USA}%

\date{\today}% It is always \today, today,

\begin{abstract}
% Motivation
Recent advances in quantum computing have drawn considerable attention to building realistic application for \textit{and} using quantum computers.
% Challenge
However, designing a suitable quantum circuit architecture requires expert knowledge. For example, it is non-trivial to design a quantum gate sequence for generating a particular quantum state with as fewer gates as possible.
% Approach
%In this paper, we 
We propose a quantum architecture search framework with the power of deep reinforcement learning (DRL) to address this challenge. In the proposed framework,  the DRL agent can only access the Pauli-$X$, $Y$, $Z$ expectation values and a predefined set of quantum operations for learning the target quantum state, and is optimized by the advantage actor-critic (A2C) and proximal policy optimization (PPO) algorithms. 
% Results & Impact
%Our proposed framework 
% Leo: it's always about us
We demonstrate a successful generation of quantum gate sequences
for multi-qubit GHZ states without encoding any knowledge of quantum physics in the agent. %In addition, the 
The design of our framework is rather general and can be employed with other DRL architectures or optimization methods to study gate synthesis and compilation for many quantum states. 

\end{abstract}

\maketitle

%\tableofcontents

\section{\label{sec:Indroduction}Introduction}
%
% \textbf{Motivation}
% \textbf{Cite relevant QML results.}
% \textbf{Cite NISQ.}
% \textbf{Cite classical deep RL results.}
% \textbf{Cite classical neural architecture search results.}

Recently, reinforcement learning (RL) \cite{sutton2018reinforcement} has found tremendous success and demonstrated a human- or superhuman- level of capabilities in a wide range of tasks, such as mastering video games \cite{Mnih2015Human-levelLearning, schrittwieser2019mastering, badia2020agent57, kapturowski2018recurrent} and even the game of Go \cite{silver2016mastering, silver2017mastering}. With such success, it is natural to consider applying such techniques to scientific areas that require sophisticated control capabilities. Indeed, RL has been used to study quantum control \cite{bukov2018reinforcement, fosel2018reinforcement, niu2019universal, an2019deep, zhang2019does, palittapongarnpim2017learning, xu2019generalizable}, quantum error correction \cite{andreasson2019quantum, fitzek2020deep, olsson2020distributed, nautrup2019optimizing, colomer2020reinforcement} and the optimization of variational quantum algorithms \cite{wauters2020reinforcement, yao2020policy, verdon2019learning, wilson2019optimizing}.
%
% \textbf{Introduce the advances of classical NAS}\\

RL has also been applied to automatically building a deep learning architecture for a given task. This is the so-called \emph{neural architecture search} \cite{zoph2016neural} and has been proven possible in a wide variety of machine learning (ML) tasks \cite{baker2016designing, cai2017efficient, zoph2018learning, zhong2018practical, schrimpf2017flexible, pham2018efficient, cai2018path}. The core idea is to train an RL agent to sequentially put in different deep learning components (e.g., convolutional operations, residual connections, pooling and so on) and then evaluate the model performance. Although the concept is simple, several recent studies have reported reaching a state-of-the-art performance \cite{elsken2019neural} and beating the best human-crafted DL models.

%
% \textbf{Introduce the promise of quantum computing}\\
Quantum computing has promised exponential speedups for several hard computational problems otherwise intractable on a classical computer \cite{harrow2017quantum, arute2019quantum}, such as factorizing large integers \cite{shor1999polynomial} and unstructured database search \cite{grover1997quantum}. Recent studies in variational quantum algorithms (VQA) have applied quantum computing to many scientific domains, including molecular dynamical studies \cite{peruzzo2014variational}, quantum optimization \cite{zhou2018quantum, farhi2014quantum} and  various quantum machine learning (QML) applications such as regression \cite{chen2020quantum, mitarai2018quantum,kyriienko2020solving}, classification \cite{mitarai2018quantum,schuld2018circuit,havlivcek2019supervised,Farhi2018ClassificationProcessors,benedetti2019parameterized,mari2019transfer, abohashima2020classification, easom2020towards, sarma2019quantum, stein2020hybrid,chen2020hybrid,chen2020qcnn,wu2020application,stein2021quclassi,chen2021hybrid,jaderberg2021quantum}, generative modeling \cite{dallaire2018quantum,stein2020qugan, zoufal2019quantum, situ2018quantum,nakaji2020quantum}, deep reinforcement learning \cite{chen19,lockwood2020reinforcement,jerbi2019quantum,Chih-ChiehCHEN2020,wu2020quantum,skolik2021quantum,jerbi2021variational}, sequence modeling \cite{chen2020quantum, bausch2020recurrent, takaki2020learning}, speech recognition \cite{yang2020decentralizing}, metric and embedding learning \cite{lloyd2020quantum, nghiem2020unified}, transfer learning \cite{mari2019transfer} and federated learning \cite{chen2021federated}.
% \textcolor{red}{(Leo: this list is too long, people would frown upon. How about breaking it into several appropriate places in the paper?)}
%
%
% \textbf{Motivate the combination of the two}\\
However, designing a quantum circuit to solve a specific task is non-trivial, as it demands domain knowledge and sometimes extraordinary insights. 

\begin{figure}[htbp]
\centering
\includegraphics[width=0.7\linewidth]{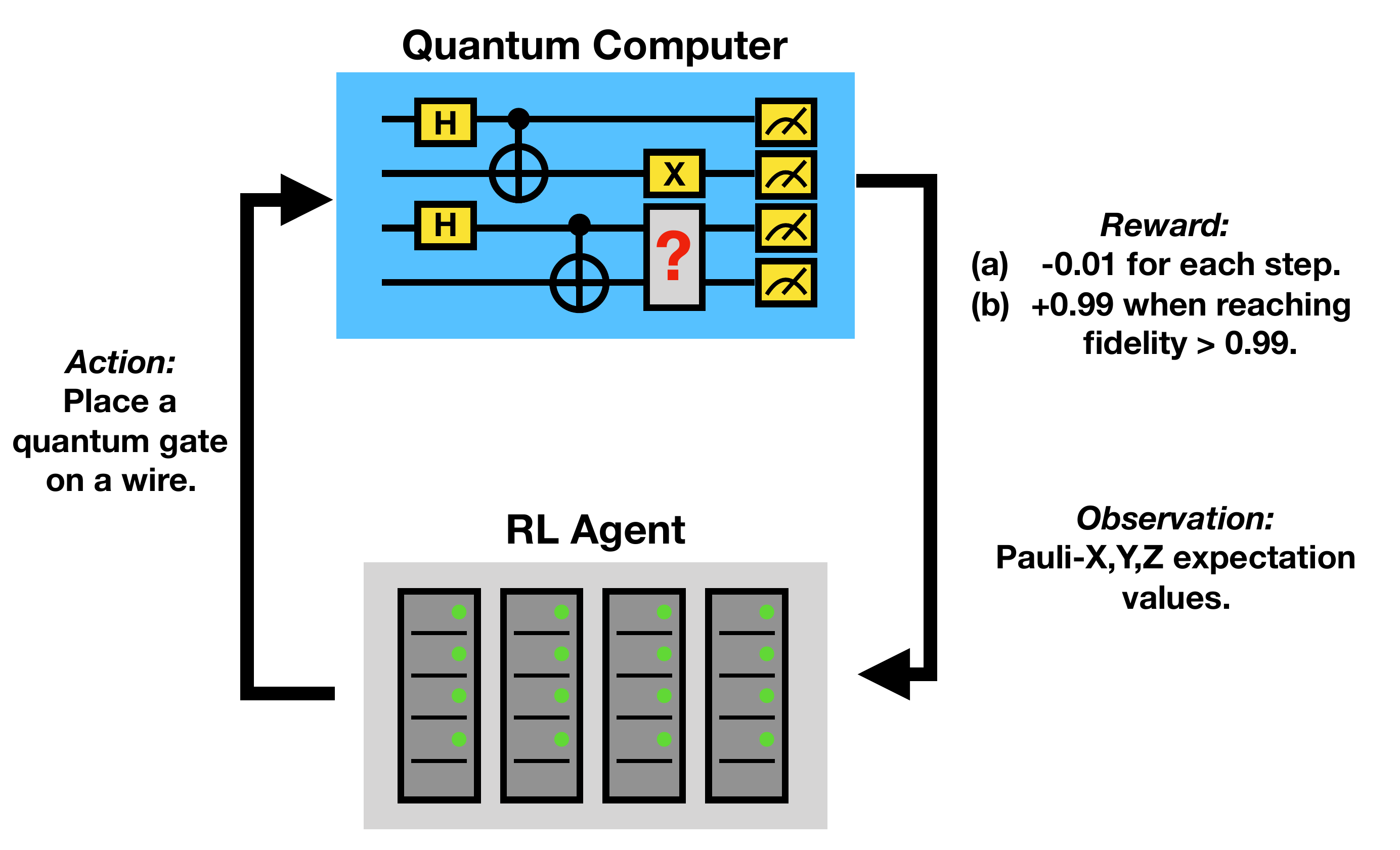}
\caption{{\bfseries Overview of DRL for our quantum architecture search framework.} The proposed quantum architecture search framework consists of two major components. First is a quantum computer or quantum simulator. In this work, we use a quantum simulator with and without noise. Second is an RL agent interacting with the quantum computer. In each time step, the RL agent will generate an action for the quantum computer. The action specifies a quantum operation to be added to the system. Then the fidelity of the quantum circuit is evaluated to determine the \emph{reward} to be sent back to the agent. In addition, Pauli-$X$, $Y$ and $Z$ expectation values are also fed back to the agent. The RL agent will then be updated based on these information. }
\label{overall_diagram}
\end{figure}

In this study, we investigate the potential of training an RL agent to search for a \emph{quantum circuit architecture} for generating a desired quantum state. 
In this work, we present a new quantum architecture search framework powered by deep reinforcement learning (DRL).
As shown in \figureautorefname{\ref{overall_diagram}}, the proposed framework includes an RL agent interacting with a quantum computer or quantum simulator. The RL agent will sequentially generate an output action, which is a candidate of the quantum gate or operation placed on the circuit. The built circuit is evaluated against certain metrics, such as the \emph{fidelity}, to check if it actually reaches the goal. The reward is calculated based on the fidelity and sent back to the RL agent. The procedure is carried out iteratively to train the RL agent.

Our contributions are the following:
\begin{itemize}
    \item Provide a framework for the study of quantum architecture search.
    \item Demonstrate building a quantum circuit step-by-step via deep reinforcement learning without any knowledge in physics.
\end{itemize}
 The paper is organized as follows. In Section~\ref{sec:ReinforcementLearning}, we introduce the RL background knowledge used in this work. In Section~\ref{sec:ProblemStatement} we introduce the quantum architectures that our agent will search. In Section~\ref{sec:ExpAndResults}, we describe the experimental procedures and results in details. Finally we discuss the results in Section~\ref{sec:Discussion} and conclude in Section~\ref{sec:Conclusion}.

\section{\label{sec:ReinforcementLearning}Reinforcement Learning}
%
% \textbf{Need to rephrase this section. Keep the contents while with different wording.....currently they are from my old articles}
%
\emph{Reinforcement learning} (RL) is a machine learning paradigm in which an \emph{agent} learns how to make decisions via interacting with the environments ~\cite{sutton2018reinforcement}. Concretely speaking, the \emph{agent} interacts with an \emph{environment} $\mathcal{E}$ over a number of discrete time steps. At each time step $t$, the agent receives a \emph{state} or \emph{observation} $s_t$ from the environment $\mathcal{E}$ and then chooses an \emph{action} $a_t$ from a set of possible actions $\mathcal{A}$ according to its \emph{policy} $\pi$. The policy $\pi$ is a function which maps the state or observation $s_t$ to action $a_t$. In general, the policy can be stochastic, meaning that given a state $s$, the action output can be a probability distribution $\pi(a_t|s_t)$ conditioned on $s_t$. After executing the action $a_t$, the agent receives the state of the next time step $s_{t+1}$ and a scalar \emph{reward} $r_t$. The process continues until the agent reaches the terminal state or a pre-defined stopping criteria (e.g. the maximum steps allowed). An \emph{episode} is defined as an agent starting from a randomly selected initial state and following the aforementioned process all the way through the terminal state or reaching a stopping criteria.

We define the total discounted return from time step $t$ as $R_t = \sum_{t'=t}^{T} \gamma^{t'-t} r_{t'}$, where $\gamma$ is the discount factor that lies in $(0,1]$. In principle, $\gamma$ is provided by the investigator to control how future rewards are weighted to the decision making function. When a large $\gamma$ is considered, the agent weighs the future reward more heavily. On the other hand, with a small $\gamma$, future rewards are quickly ignored and immediate reward will be weighted more. 
The goal of the agent is to maximize the expected return from each state $s_t$ in the training process. The \emph{action-value function} or \emph{Q-value function} $Q^\pi (s,a) = \mathbb{E}[R_t|s_t = s, a]$ is the expected return for selecting an action $a$ in state $s$ based on policy $\pi$. The optimal action value function $Q^*(s,a) = \max_{\pi} Q^\pi(s,a)$ gives a maximal action-value across all possible policies. The value of state $s$ under policy $\pi$, $V^\pi(s) = \mathbb{E}\left[R_t|s_t = s\right]$, is the agent's expected return by following policy $\pi$ from the state $s$. Various RL algorithms are designed to find the policy which can maximize the value function. The RL algorithms which maximize the value function are called \emph{value-based} RL. 
\subsection{Policy Gradient}
In contrast to the \emph{value-based} RL, which learns the value function and use it as the reference to generate the decision on each time-step, there is another kind of RL method called \emph{policy gradient}. In this method, the policy function $\pi(a|s;\theta)$ is parameterized with the  parameters $\theta$. The $\theta$ will then be subject to the optimization procedure which is \emph{gradient ascent} on the expected total return $\mathbb{E}[R_{t}]$. One of the classic examples of policy gradient algorithm is the REINFORCE algorithm~\cite{williams1992simple}. In the standard REINFORCE algorithm, the parameters $\theta$ are updated along the direction $\nabla_{\theta} \log \pi\left(a_{t} | s_{t} ; \theta\right) R_{t}$, which is the unbiased estimate of $\nabla_{\theta} \mathbb{E}\left[R_{t}\right]$. However, the policy gradient method suffers from large variance of the $\nabla_{\theta} \mathbb{E}\left[R_{t}\right]$, making the training very hard. To reduce the variance of this estimate and keep it unbiased, one can subtract a learned function of the state $b_{t}(s_{t})$, which is known as the \emph{baseline}, from the return. The result is therefore $\nabla_{\theta} \log \pi\left(a_{t} | s_{t} ; \theta\right)\left(R_{t}-b_{t}\left(s_{t}\right)\right)$.

\subsection{Advantage Actor-Critic (A2C)}
A learned estimate of the value function is a common choice for the baseline $b_{t}(s_{t}) \approx V^{\pi}(s_{t})$. This choice usually leads to a much lower variance estimate of the policy gradient. When one uses the approximate value function as the baseline, the quantity $R_{t} - b_{t} = Q(s_{t}, a_{t}) - V(s_{t})$ can be seen as the \emph{advantage} $A(s_{t}, a_{t})$ of the action $a_{t}$ at the state $s_{t}$. Intuitively, one can see this advantage as ``how good or bad the action $a_{t}$ compared to the average value at this state $V(s_{t})$.'' For example, if the $Q(s_{t}, a_{t})$ equals to $10$ at a given time-step $t$, it is not clear whether $a_{t}$ is a good action or not. However, if we also know that the $V(s_{t})$ equals to, say $2$ here, then we can imply that $a_{t}$ may not be bad. Conversely, if the $V(s_{t})$ equals to $15$, then the advantage is $10 - 15 = -5$, meaning that the $Q$ value for this action $a_{t}$ is well below the average $V(s_{t})$ and therefore that action is not good. This approach is called \emph{advantage actor-critic} (A2C) method where the policy $\pi$ is the actor and the baseline which is the value function $V$ is the critic~\cite{sutton2018reinforcement}.
\subsection{Proximal Policy Optimization (PPO)}
In the policy gradients method, we optimize the policy according to the \emph{policy loss} $L_{\text{policy}}(\theta) = \mathbb{E}_{t}[-\log\pi\left(a_{t}\mid s{t};\theta \right)]$ via gradient descent. However, the training itself may suffer from instabilities. If the step size of policy update is too small, the training process would be too slow. On the other hand, if the step size is too high, there will be a high variance in the training. The proximal policy optimization (PPO) \cite{schulman2017proximal} fixes this problem by limiting the policy update step size at each training step. The PPO introduces the loss function called \emph{clipped surrogate loss function} that will constraint the policy change a a small range with the help of a clip. Consider the ratio between the probability of action $a_{t}$ under current policy and the probability under previous policy  $q_{t}(\theta)=\frac{\pi\left(a_{t} \mid s_{t} ; \theta \right)}{\pi\left(a_{t} \mid s_{t} ; \theta_{\text{old}}\right)}$. If $q_{t}(\theta) > 1$, it means the action $a_{t}$ is with higher probability in the current policy than in the old one. And if $0 < q_{t}(\theta) < 1$, it means that the action $a_{t}$ is less probable in the current policy than in the old one. Our new loss function can then be defined as $L_{\text{policy}}(\theta) = \mathbb{E}_{t}[q_{t}(\theta)  A_{t}] = \mathbb{E}_{t}[\frac{\pi\left(a_{t} \mid s_{t} ; \theta \right)}{\pi\left(a_{t} \mid s_{t} ; \theta_{\text{old}}\right)}  A_{t}]$, where $A_{t} = R_{t} - V(s_{t}; \theta)$ is the advantage function. However, if the action under current policy is much more probable than in the previous policy, the ratio $q_{t}$ may be large, leading to a large policy update step. To circumvent this problem, the original PPO algorithm \cite{schulman2017proximal} adds a constraint on the ratio, which can only be in the range $0.8$ to $1.2$. The modified loss function is now $L_{\text{policy}}(\theta) = \mathbb{E}_{t}[-min(q_{t}A_{t}, clip(q_{t}, 1 - C, 1 + C)A_{t})]$ where the $C$ is the clip hyperparameter (common choice is $0.2$). Finally, the value loss and entropy bonus are added into the total loss function as usual: $L(\theta) = L_{\text{policy}} + c_1 L_{\text{value}} - c_2 H$ where $L_{\text{value}} = \mathbb{E}_{t}[\norm{R_{t} - V(s_{t} ; \theta)}^2]$ is the value loss and $H = \mathbb{E}_{t}[H_{t}] = \mathbb{E}_{t}[-\sum_{j}\pi\left(a_{j} \mid s_{t} ; \theta\right) \log(\pi\left(a_{j} \mid s_{t} ; \theta\right))]$ is the entropy bonus which is to encourage exploration.
%
% \subsection{Recurrent Policy}
% %
% In training a deep reinforcement learning agent, occasionally the environment need the agent to have the capabilities to memorize previous observations or states in order to generate appropriate or good actions in later time-steps. 
%
\section{\label{sec:ProblemStatement}Problem Setup}
%
% \textbf{Add high level description of the problem}

Below we describe in detail the problem we aim to solve using DRL. Given an initial state $\ket{0\cdots0}$ and the target state, the goal is to produce a quantum circuit which transforms the initial state to the target state within certain error tolerance. We use the Pauli measurements as observations, a natural choice in quantum mechanics. We then use various RL algorithms to achieve our goal. The overall scheme is shown in \figureautorefname{\ref{overall_diagram}}. 
Specifically, the environment $\mathcal{E}$ is the quantum computer or quantum simulator. In this work, we use a quantum simulator since currently it is not yet practical to train tens of thousands of episodes on a cloud-based quantum device. The RL agent, hosted on a classical computer, interacts with the environment $\mathcal{E}$. 
% As mentioned in previous section on RL (see \sectionautorefname{\ref{sec:ReinforcementLearning}}),
In each time step, the RL agent  chooses an action $a$ from the possible set of actions $\mathcal{A}$, which consists of different quantum operations (one- and two- qubit gates).
After the RL agent updates the quantum circuit with the chosen action, the environment $\mathcal{E}$  executes the new circuit and calculates the \emph{fidelity} to the given target state. If the fidelity reaches a pre-defined threshold, the episode ends and a large positive reward is given to the RL agent. Otherwise, the RL agent receives a small negative reward.
The states or observations which the environment $\mathcal{E}$ returns to the RL agent are Pauli measurements on each qubit, so for an $n$-qubit system the dimension of the observations is  $3n$.
The procedure continues until the agent reaches either the desired threshold or the maximum allowed steps. RL algorithms like A2C and PPO are employed to optimize the agent.
Next, we discuss in detail the mathematical setting of our problem.

\subsection{Mathematical formulation of the problem}
%We first set up a mathematical formulation of the problem. 
Suppose we are given the number of qubits $n \in \mathbb{N}$,  the initial quantum state $|0\rangle ^{\otimes^n}$, 
the target state $|\psi\rangle$, the tolerance error $ \epsilon \geq 0$, and a set of gates $\mathbb{G}$.
Our goal is to find a quantum circuit $\mathcal{C}:|0\rangle ^{\otimes^n} \to |\psi\rangle$ so that our DRL architecture serves as a function $\mathcal{F}$: 
\begin{equation}
\mathcal{F}: (|0\rangle ^{\otimes^n},|\psi\rangle, \epsilon,\mathbb{G})\to \mathcal{C} 
\end{equation}
such that $1 \geq D(|\psi\rangle,\mathcal{C}(|0\rangle ^{\otimes^n})) \geq 1-\epsilon$, where $\mathcal{C}$ is composed of gates $g \in \mathbb{G}$ and $D$ is a distance metric between two quantum states (larger is better). In this paper, we use the fidelity \citep{nielsen2002quantum} to be our distance $D$. Given two density operators $\rho$  and $\sigma$ (see also Sec.~\ref{sec:density matrix}), the fidelity is generally defined as the quantity $ {\displaystyle F(\rho ,\sigma )=\left[\operatorname {tr} {\sqrt {{\sqrt {\rho }}\sigma {\sqrt {\rho }}}}\right]^{2}}$. In the special case where $\rho$  and $\sigma$  represent pure quantum states, namely, ${\displaystyle \rho =|\psi _{\rho }\rangle \!\langle \psi _{\rho }|}$ and ${\displaystyle \sigma =|\psi _{\sigma }\rangle \!\langle \psi _{\sigma }|}$, the definition becomes the inner product of two states: ${\displaystyle F(\rho ,\sigma )=|\langle \psi _{\rho }|\psi _{\sigma }\rangle |^{2}}$.

\subsection{Multi-qubit entangled states as target}
To validate that the proposed DRL pipeline can be applied to quantum architecture search, it is best to check if multi-qubit entanglement can be generated as expected. To this end, we target the generation of two kinds of quantum states: Bell state and Greenberger–Horne–Zeilinger (GHZ) state. 

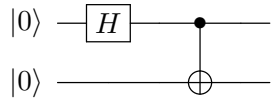
\begin{figure}[!htbp]
\begin{center}
\begin{minipage}{10cm}
\Qcircuit @C=1em @R=1em {
\lstick{\ket{0}} & \gate{H}  & \qw        & \ctrl{1}       & \qw      & \qw \\
\lstick{\ket{0}} & \qw       & \qw        & \targ          & \qw      & \qw
}
\end{minipage}
\end{center}
\caption[Quantum Circuit for Bell state.]{{\bfseries Quantum circuit for the Bell state.}
}
\label{Fig:circuitForBellState}
\end{figure}

A Bell state reaches maximal two-qubit entanglement,
\begin{equation}
\ket{\mathrm{Bell}} = \frac{\ket{0}^{\otimes 2}+\ket{1}^{\otimes 2}}{\sqrt{2}} = \frac{\ket{00}+\ket{11}}{\sqrt{2}}.
\end{equation}
To generate a Bell state, we pick the observation to be the expectation values of Pauli matrices on each qubits $\{\langle \sigma^{i}_{j}\rangle \,|\, i\in\{0, 1\}, j\in\{x, y, z\}\}$. The action set $\mathbb{G}$ is 
\begin{equation}
\mathbb{G}=\bigcup \limits_{i=1}^{n}  \left\{U_i\left(\pi/4\right), X_i, Y_i, Z_i, H_i, CNOT_{i,(i+1)(mod2)}\right\},
\label{eq:action gates}
\end{equation}
where $n=2$ (for two qubits), $U_i(\theta)=\big(\begin{smallmatrix}
  1 & 0\\
  0 & \exp(i\theta)
\end{smallmatrix}\big)$  is the single qubit rotation about the $Z$-axis applied to qubit $i$,
$X_i\equiv\sigma^i_x$ is the Pauli-$X$ gate and likewise for $Y_i$ and $Z_i$,
$H_i$ is the Hadamard gate,
and $CNOT_{i, j}$ is the CNOT gate with the $i$-th qubit as control and $j$-th qubit as target,
%\textcolor{red}{(Leo: check if this description is consistent with your notation)},\textcolor{green}{en jui: for each $x,y,z$ we use $\frac{\pi}{4}$ rotation}
so we have $12$ actions in total. %\textcolor{red}{(Leo: I think it's 14? We didn't count identity gates)} \textcolor{green}{en jui: yes, we did not. $12= 6 \times 2.$}%The target is the maximal entanglement state, i.e standard bell state.
A textbook example for creating a Bell state is shown in Fig.~\ref{Fig:circuitForBellState}.

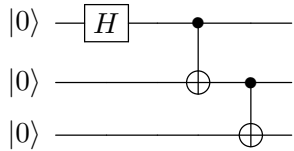
\begin{figure}[!htbp]
\begin{center}
\begin{minipage}{10cm}
\Qcircuit @C=1em @R=1em {
\lstick{\ket{0}} & \gate{H}  & \qw        & \ctrl{1}    & \qw      & \qw \\
\lstick{\ket{0}} & \qw       & \qw        & \targ       & \ctrl{1} & \qw \\
\lstick{\ket{0}} & \qw       & \qw        & \qw         & \targ      & \qw
}
\end{minipage}
\end{center}
\caption[Quantum Circuit for GHZ state.]{{\bfseries Quantum circuit for the GHZ state.}
}
\label{Fig:circuitForGHZState}
\end{figure}

%
%\subsection{Greenberger–Horne–Zeilinger (GHZ) state}
%
A GHZ state is a multi-qubit generalization of the Bell state, in which an equal superposition between the lowest and the highest energy states is created. For 3 qubits it is given by
\begin{equation}
\ket{\mathrm{GHZ}} = \frac{\ket{0}^{\otimes 3}+\ket{1}^{\otimes 3}}{\sqrt{2}} = \frac{\ket{00 0}+\ket{111}}{\sqrt{2}}
\end{equation}
To generate the 3-qubit GHZ state, we again use the expectation values of individual qubit's Pauli matrices, leading to $9$ observables in total. For the actions, we pick %$U(\frac{\pi}{4})_i, X_i, Y_i, Z_i,H_i$ as single qubit gates 
the same single-qubit gates as in Eq.~\eqref{eq:action gates},
and six $CNOT$ gates %\textcolor{red}{(Leo: which three?)} \textcolor{green}{enjui: Thanks for pointing out, there should be $6$ of them, $(12), (13), (23)$ and $(21), (31), (32)$ } 
as two-qubit gates, so total we have $21$ actions. 
%\textcolor{red}{(Leo: from the old description I counted 18, have we included identity gates here?)} \textcolor{green}{enjui: as I pointed here, for we have $3 \times 5+6=21$. We do not count identities.} 
In this fashion, for general $n$-qubit cases there will be $5n+n(n-1)= \Omega(n^2)$ actions, 
%($5n$ comes from 
%the same single-qubit gates as in Eq.~\eqref{eq:action gates}, and $n(n-1)$ comes from $n$ choose $2$ times $2$ $CNOT$ gates), 
increasing only quadratically instead of exponentially in $n$. 
%%The target what we use to demonstrate is the GHZ state described below.
% %
% \iffalse
% \begin{equation}
% \ket{\mathrm{GHZ}} = \frac{\ket{0}^{\otimes M}+\ket{1}^{\otimes M}}{\sqrt{2}} = \frac{\ket{0 \cdots 0}+\ket{1 \cdots 1}}{\sqrt{2}}
% \end{equation}
% \fi
% %
An example for creating a 3-qubit GHZ state is shown in Fig.~\ref{Fig:circuitForGHZState}.

\section{\label{sec:ExpAndResults}Experiments and Results}
%
% \textbf{need to fill the algorithm boxes}
% \textbf{need to check the A2C in detail on the loss calculation and updates}
%

%
\subsection{Experimental Settings}

\subsubsection{Optimizer}
We apply the gradient-descent method to optimize the RL policy. There are a wide variety of gradient-descent methods which are demonstrated highly successful \cite{ruder2016overview, Tieleman2012, kingma2014adam}. 
In this work, we use the Adam \cite{kingma2014adam} optimizer for training the RL agent in both the A2C and PPO cases. Adam is one of the gradient-descent methods which computes the \emph{adaptive learning rates} for each parameter. In addition, Adam stores both the exponentially decaying average of gradients $g_t$ and its square $g_{t}^2$,
\begin{subequations}
\begin{align} 
m_{t} &=\beta_{1} m_{t-1}+\left(1-\beta_{1}\right) g_{t} \\ 
v_{t} &=\beta_{2} v_{t-1}+\left(1-\beta_{2}\right) g_{t}^{2} 
\end{align}
\end{subequations}
where $\beta_1$ and $\beta_2$ are hyperparameters. We use $\beta_{1} = 0.9$ and $\beta_{2} = 0.999$ in this work.
The $m_t$ and $v_t$ are adjusted according to the following formula to counteract the biases towards $0$,
\begin{subequations}
\begin{align} 
\hat{m}_{t} &=\frac{m_{t}}{1-\beta_{1}^{t}} \\ 
\hat{v}_{t} &=\frac{v_{t}}{1-\beta_{2}^{t}} \end{align}
\end{subequations}
The parameters $\theta_{t}$ in the RL model in the time step $t$ are then updated according to the following formula,
\begin{equation}
\theta_{t+1}=\theta_{t}-\frac{\eta}{\sqrt{\hat{v}_{t}}+\epsilon} \hat{m}_{t}
\end{equation}
We use the Adam optimizer provided in the Python package \textsc{PyTorch} \cite{NEURIPS2019_9015_PYTORCH} to perform the optimization procedures.

\subsubsection{Quantum Noise in Quantum Simulator}
% \textbf{Introduce the concept of gate error and measurement error}
%
Here we introduce the error schemes we use in this study. We consider two forms of errors, \emph{gate errors} and  \emph{measurement errors}.
The gate error refers to the imperfection in any quantum operation we perform, whereas the measurement error refers to the error that occurs during quantum measurement. For the gate error, we consider the depolarizing noise which replaces the state of any qubit with a random state of probability $p_{gate}$. For the measurement error, we consider a random flip between $0$ and $1$ with probability $p_{meas}$ immediately before the actual measurement.
We use the following noise configuration in the simulation software to test our deep RL agents:
\begin{itemize}
    \item error rate (both $p_{gate}$ and $p_{meas}$) $= 0.001$
    \item error rate (both $p_{gate}$ and $p_{meas}$) $= 0.005$
\end{itemize}
%
%Our noisy quantum simulation is implemented with the Qiskit \cite{cross2018ibm} software package.
For the simulation of quantum circuits in both noise-free and noisy environments, we use the software package Qiskit from IBM \cite{cross2018ibm}.

\subsubsection{Density Matrix of Quantum States}
\label{sec:density matrix}
The general form of a \emph{density matrix} $\rho$ of a quantum state under the basis $\{\ket{\psi_i}\}$ is, 
\begin{equation}
\rho=\sum_{j} p_{j}\ket{\psi_{j}}\bra{\psi_{j}}
\end{equation}
where $p_j$ represents the probability that the quantum system is in the pure state $\ket{\psi_{j}}$ such that $\sum_{j} p_{j} = 1$. For example, the density matrix of the Bell state considered in this study is $\ket{\mathrm{Bell}} = \left(\ket{00}+\ket{11}\right)/\sqrt{2}$. Its corresponding density matrix $\rho$ is then given by
\begin{equation}
    \ket{\mathrm{Bell}}\bra{\mathrm{Bell}} = \frac{1}{2}\left(\ket{00}\bra{00}+\ket{00}\bra{11}+\ket{11}\bra{00}+\ket{11}\bra{11}\right)
\end{equation}
The density matrix is used in calculating the state fidelity $F$ as mentioned earlier.

\subsubsection{Quantum State Tomography}
Quantum state tomography is a procedure to reconstruct the density matrix associated with a quantum state from a set of complete measurements. Expanding the density matrix in the Pauli basis of $N$ qubits,
\begin{equation}
    \rho = \frac{1}{2^N} \sum_{i_1, \cdots, i_N = 0}^3 \rho_{i_i, \cdots, i_N} \sigma_{i_1} \otimes \cdots \otimes \sigma_{i_N},
    \label{eq:QST}
\end{equation}
it can be seen that to fully determine $\rho$ requires $4^N-1$ measurement operations (minus one due to the conservation of probability, $\text{Tr}(\rho)=1$).
More generally, measurements with $4^N-1$ linearly independent projective operators can uniquely determine the density matrix, for which Eq.~\ref{eq:QST} is a special case with the projectors being the Pauli operators.
As a result, the number of measurements grows exponentially in the qubit number $N$, posing a significant challenge in verifying multi-qubit quantum states in any experiments, and with a finite number of shots the expectation values for $\{\rho_{i_1, \cdots, i_N}\}$ can only be measured within certain accuracy.
For the purpose of this work, however, we perform the quantum state tomography simulations using IBM's Qiskit software package \cite{cross2018ibm}.

\subsubsection{Customized OpenAI Gym Environment}
We build a customized OpenAI Gym \cite{brockman2016openai} environment to facilitate the development and testing of this work. In this package, users can set the target quantum state, threshold of fidelity and the quantum computing backend (real device or simulator software). In addition, it is also possible to customize the noise pattern.  
% Here we present the configuration of the testing environment.
% \YC{Put the reward scheme here}
We construct the testing environments with the following settings:
%
% \YC{env setting? like the tolerance and num of qubits?}
\begin{itemize}
    \item \textbf{Observation:} As mentioned the agent receives Pauli-$X$, $Y$ $Z$ expectation values on each qubit. %For example, in the $2$-qubit case, the dimension of the observation is $3 \times 2 = 6$ and in the $3$-qubit case it is then $3 \times 3 = 9$. 
    For general $n$-qubit systems, the number of observations will be $3 \times n$.
    \item \textbf{Action:} The RL agent is expected to select a quantum gate operating on the specific qubit as given in Eq.~\eqref{eq:action gates}. %We define the possible gate set to include single-qubit gates on each qubit and two-qubit gate $CNOT$ on each pair of qubits. In the $2$-qubit case, the action space is $\bigcup \limits_{i=1}^{2}  \{U(\frac{\pi}{4})_i, X_i, Y_i, Z_i,H_i, CNOT_{i,(i+1)(mod2)}\}$. The number of actions for $2$-qubit system is $12$. In the $3$-qubit case, the total number of actions is $21$. For general $n$-qubit system, the number of actions will be $5n+\frac{n(n-1)}{2}$.
    \item \textbf{Reward:} For each step before successfully reaching the goal, the agent will receive a $-0.01$ reward to encourage the shortest path. When reaching the goal, the agent will receive a reward of value ($F - 0.01$).
\end{itemize}
\subsubsection{Hyperparameters}
In this work, we employ the neural network models (shown in \tableautorefname{\ref{tab:nn_a2c_ppo}}) as our DRL agents:
% \textbf{Write the neural network structure here.}
% \YC{go back to this later}
%

%
We consider two DRL algorithms in this work, their hyperparameters are:
\begin{itemize}
    \item A2C: learning rate $\eta = 10^{-4}$, discount factor $\gamma = 0.99$
    \item PPO: learning rate $\eta = 0.002$, discount factor $\gamma = 0.99$, epsilon clip parameter $C = 0.2$, update epoch number $K = 4$
\end{itemize}
\begin{table}[htbp]
\begin{tabular}{|l|l|l|l|l|l|l|l|}
\hline
       & Linear    & \multicolumn{2}{l|}{Tanh}              & Linear & \multicolumn{2}{l|}{Tanh}              & Linear                           \\ \hline
Input  & state dim & \multicolumn{2}{l|}{\multirow{2}{*}{}} & 64     & \multicolumn{2}{l|}{\multirow{2}{*}{}} & 64                               \\ \cline{1-2} \cline{5-5} \cline{8-8} 
Output & 64        & \multicolumn{2}{l|}{}                  & 64     & \multicolumn{2}{l|}{}                  & action dim (actor) or 1 (critic) \\ \hline
\end{tabular}
\caption{The neural network for A2C and PPO. The structure is the same for both the actor and critic. The only difference is that in the actor, there is a softmax at the end of the network.}
\label{tab:nn_a2c_ppo}
\end{table}

\subsection{Results: Noise-Free Environments}
\subsubsection{$2$-qubit Bell state}
%
% \textbf{place an example circuit compared with standard}
Here we consider the application of DRL to generate the $2$-qubit Bell state from scratch under the noise-free environment. 
The result is in the \figureautorefname{\ref{fig:noise_free_two_qubit}}.
We can observe that both A2C and PPO methods can successfully train the DRL agent to synthesize the Bell state. It is demonstrated that, with the same neural network architecture, the PPO method reaches optimal results faster and the result is more stable compared to the A2C method. In \figureautorefname{\ref{Fig:circuitForBellState_from_PPO}} we provide the quantum circuit for Bell state generated by the DRL agent. 
\begin{figure}[!htbp]
\begin{center}
\begin{minipage}{10cm}
\Qcircuit @C=1em @R=1em {
\lstick{\ket{0}} & \qw            & \qw        & \targ       & \qw      & \qw \\
\lstick{\ket{0}} & \gate{H}       & \qw        & \ctrl{-1}   & \qw      & \qw
}
\end{minipage}
\end{center}
\caption[Quantum Circuit for Bell state.]{{\bfseries Quantum circuit for the Bell state generated by the DRL(PPO) agent.}
}
\label{Fig:circuitForBellState_from_PPO}
\end{figure}
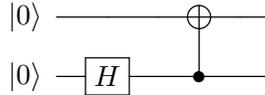
\begin{figure}[htbp]
     \begin{subfigure}[t]{1\textwidth}
     \centering
         \scalebox{0.5}{
            \includegraphics{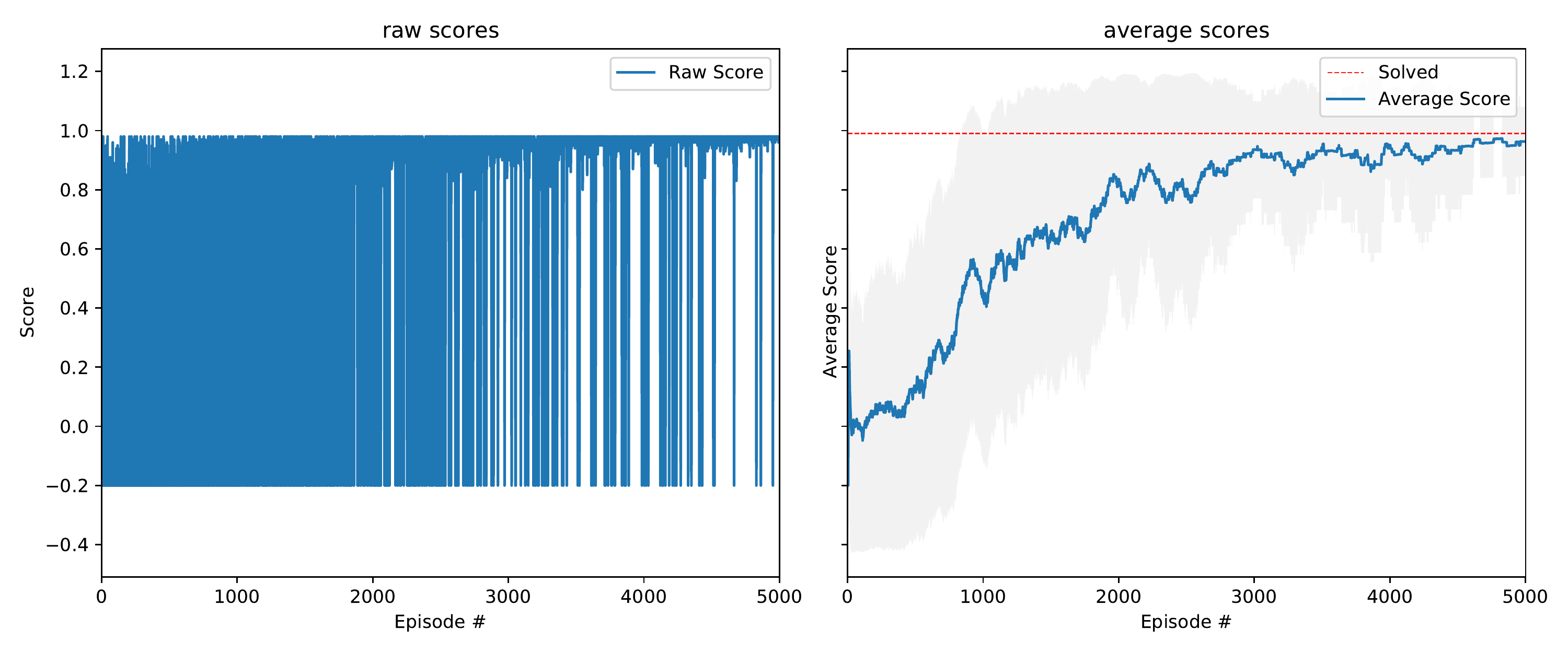}% Here is how to import EPS art
            }
        \caption{\label{fig:noise_free_two_qubits_a2c} {\bfseries A2C for noise-free two-qubit system.}
        }
     \end{subfigure}
     \hfill
     \begin{subfigure}[t]{1\textwidth}
     \centering
         \scalebox{0.5}{
            \includegraphics{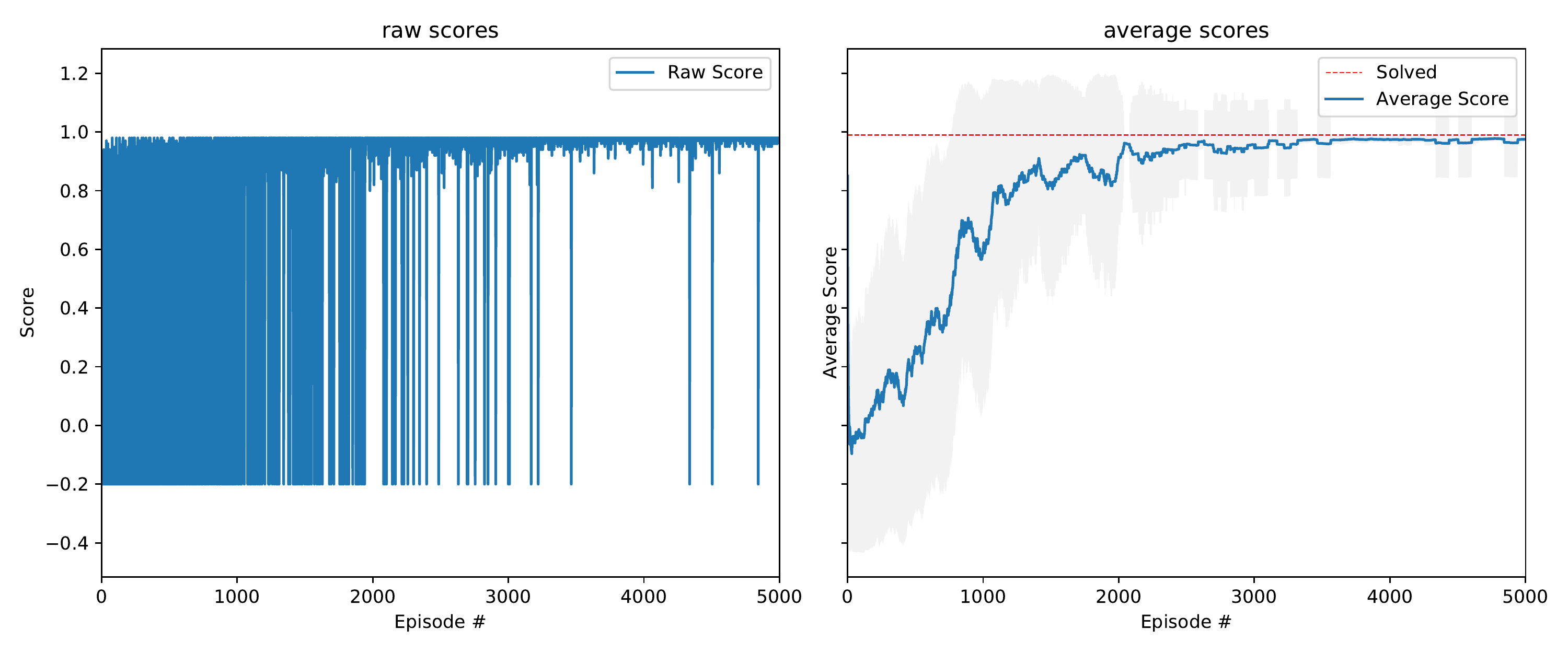}% Here is how to import EPS art
            }
        \caption{\label{fig:noise_free_two_qubits_ppo} {\bfseries PPO for noise-free two-qubit system.}}
     \end{subfigure}
     \hfill
    \caption{{\bfseries Deep Reinforcement Learning for Two-Qubit system.} In the synthesis of the Bell state with noise-free simulation environment, we set the total number of training episodes to be $5000$. The left panels of the figure show the raw scores of the DRL agents. The gray area in the right panels of the figure represents the standard deviation of reward in each training episode. We observe that, given the same neural network architecture, PPO performs better than the A2C in terms of the convergence speed and the stability. }
    \label{fig:noise_free_two_qubit}
\end{figure}

\subsubsection{$3$-qubit GHZ state}
Here we consider the application of DRL to generate the $3$-qubit GHZ state from scratch under the noise-free environment.
The result is in the \figureautorefname{\ref{fig:noise_free_three_qubit}}. We can observe that both A2C and PPO methods can successfully train the DRL agent to synthesize the GHZ state. It is demonstrated that, with the same neural network architecture, the PPO method reaches optimal results faster and the result is more stable compared to the A2C method. Notably, the advantage of PPO over A2C is much more significant compared to the $2$-qubit case. In \figureautorefname{\ref{Fig:circuitForGHZState_from_PPO}} we provide the quantum circuit for GHZ state generated by the DRL agent. 
\begin{figure}[!htbp]
\begin{center}
\begin{minipage}{10cm}
\Qcircuit @C=1em @R=1em {
\lstick{\ket{0}} & \qw.      & \qw        & \qw.        & \targ     & \qw \\
\lstick{\ket{0}} & \gate{H}  & \qw        & \ctrl{1}    & \ctrl{-1} & \qw \\
\lstick{\ket{0}} & \qw       & \qw        & \targ       & \qw       & \qw
}
\end{minipage}
\end{center}
\caption[Quantum Circuit for GHZ state.]{{\bfseries Quantum circuit for the GHZ state generated by the DRL(PPO) agent.}
}
\label{Fig:circuitForGHZState_from_PPO}
\end{figure}
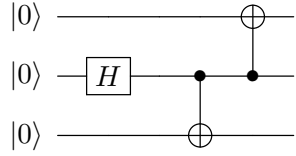
\begin{figure}[htbp]
     \begin{subfigure}[t]{1\textwidth}
     \centering
         \scalebox{0.5}{
            \includegraphics{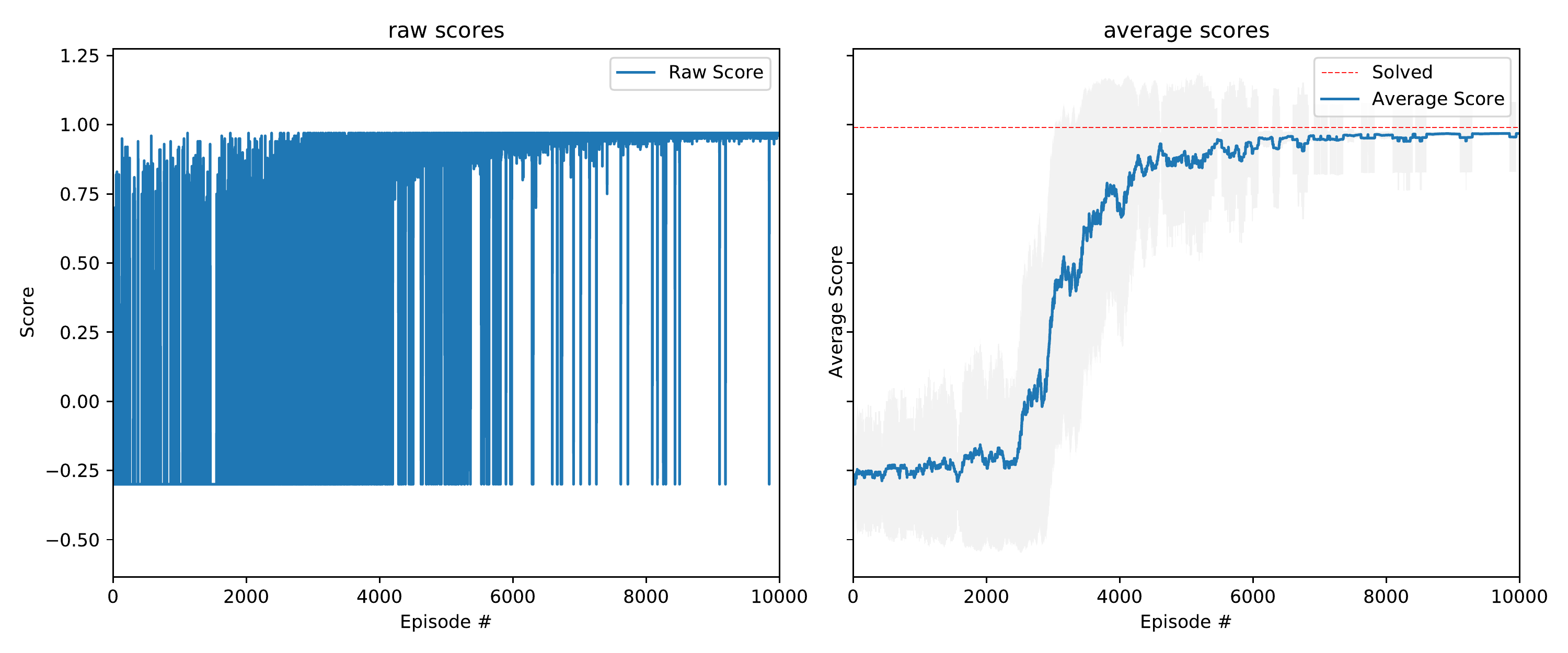}% Here is how to import EPS art
            }
        \caption{\label{fig:noise_free_three_qubits_a2c} {\bfseries A2C for noise-free three-qubit system.}
        }
     \end{subfigure}
     \hfill
     \begin{subfigure}[t]{1\textwidth}
     \centering
         \scalebox{0.5}{
            \includegraphics{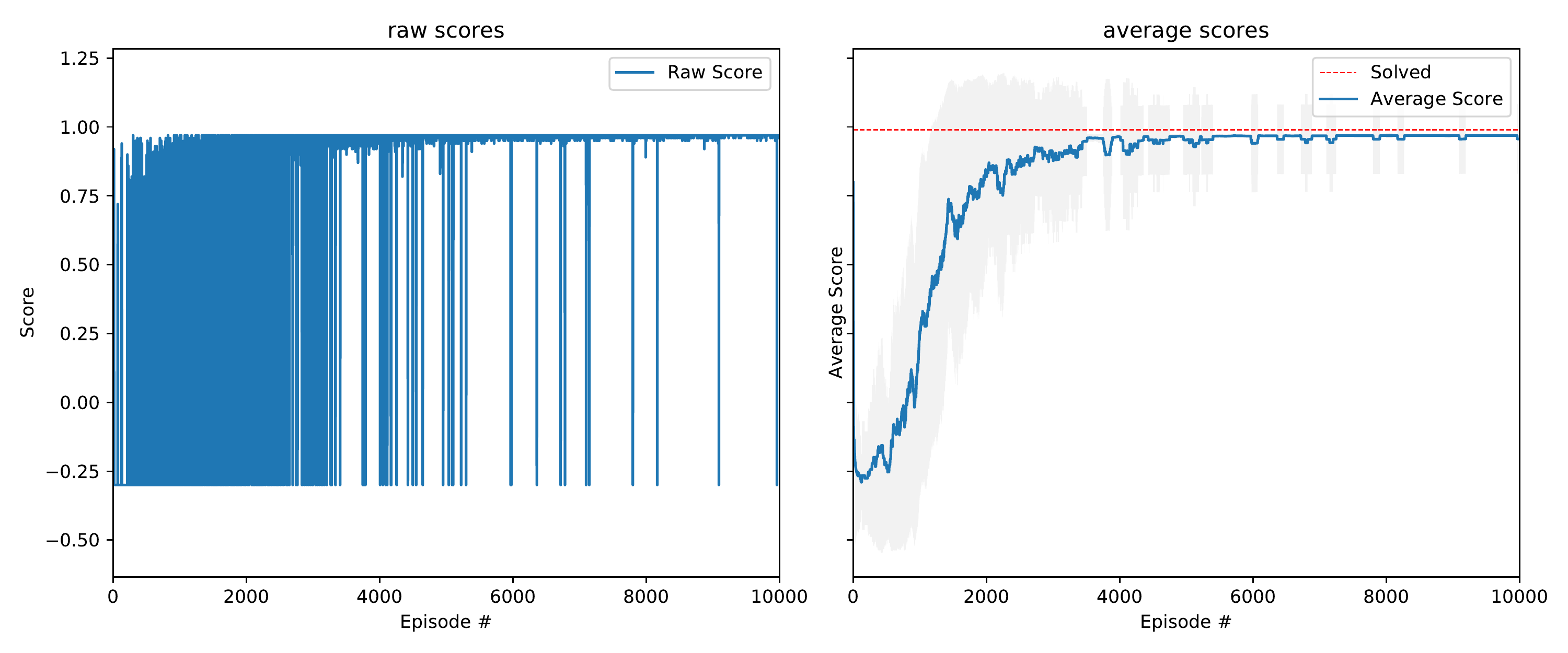}% Here is how to import EPS art
            }
        \caption{\label{fig:noise_free_three_qubits_ppo} {\bfseries PPO for noise-free three-qubit system.}}
     \end{subfigure}
     \hfill
    \caption{{\bfseries Deep Reinforcement Learning for Three-Qubit system.} In the synthesis of the GHZ state with noise-free simulation environment, we set the total number of training episodes to be $10000$. We observe that, given the same neural network architecture, PPO performs significantly better than the A2C in terms of the convergence speed and the stability. The result is consistent with the $2$-qubit case.}
    \label{fig:noise_free_three_qubit}
\end{figure}

\subsection{Results: Noisy Environments}
\begin{figure}[htbp]
     \begin{subfigure}[t]{1\textwidth}
     \centering
         \scalebox{0.4}{
            \includegraphics{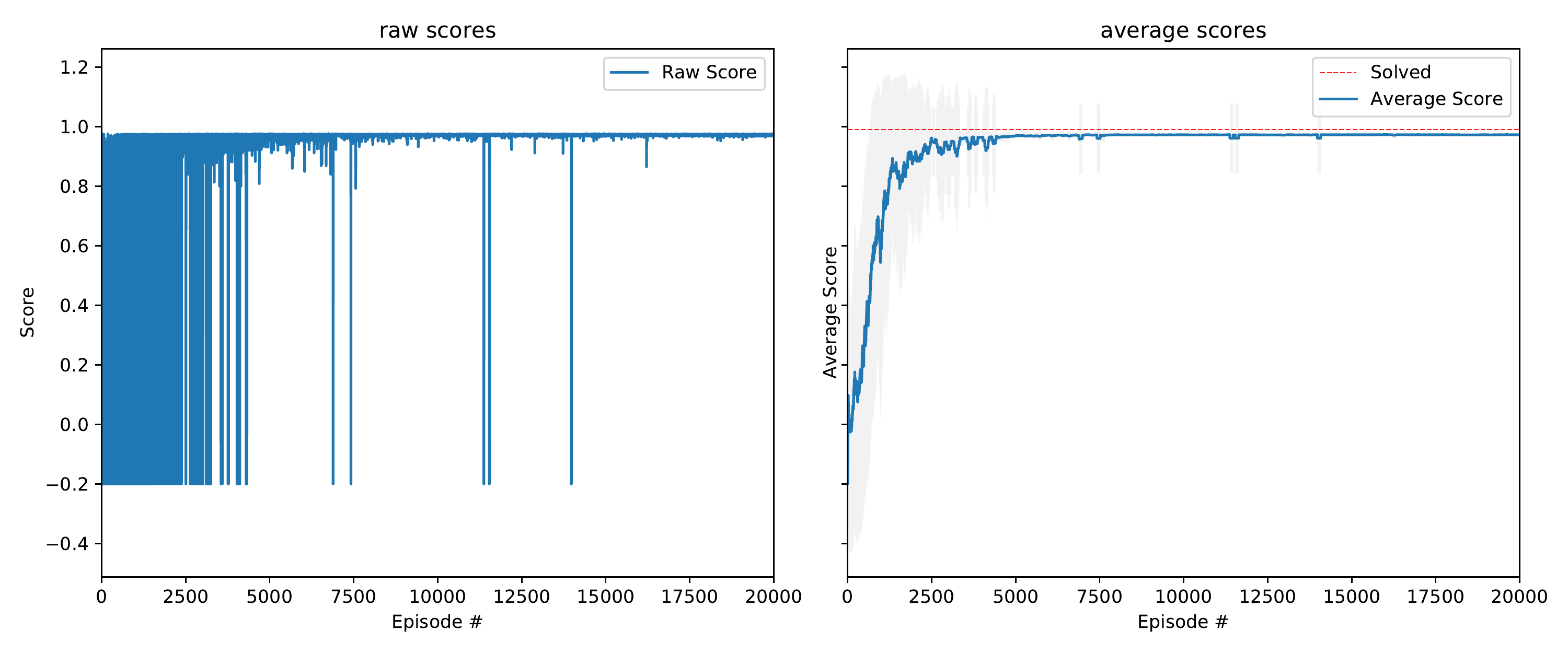}% Here is how to import EPS art
            }
        \caption{\label{fig:noisy_two_qubits_ppo_noise0001_fidelity_threshold_095} {\bfseries PPO for noisy two-qubit system with single-qubit error rate $0.001$ and fidelity threshold $0.95$.}
        }
     \end{subfigure}
     \hfill
     \begin{subfigure}[t]{1\textwidth}
     \centering
         \scalebox{0.4}{
            \includegraphics{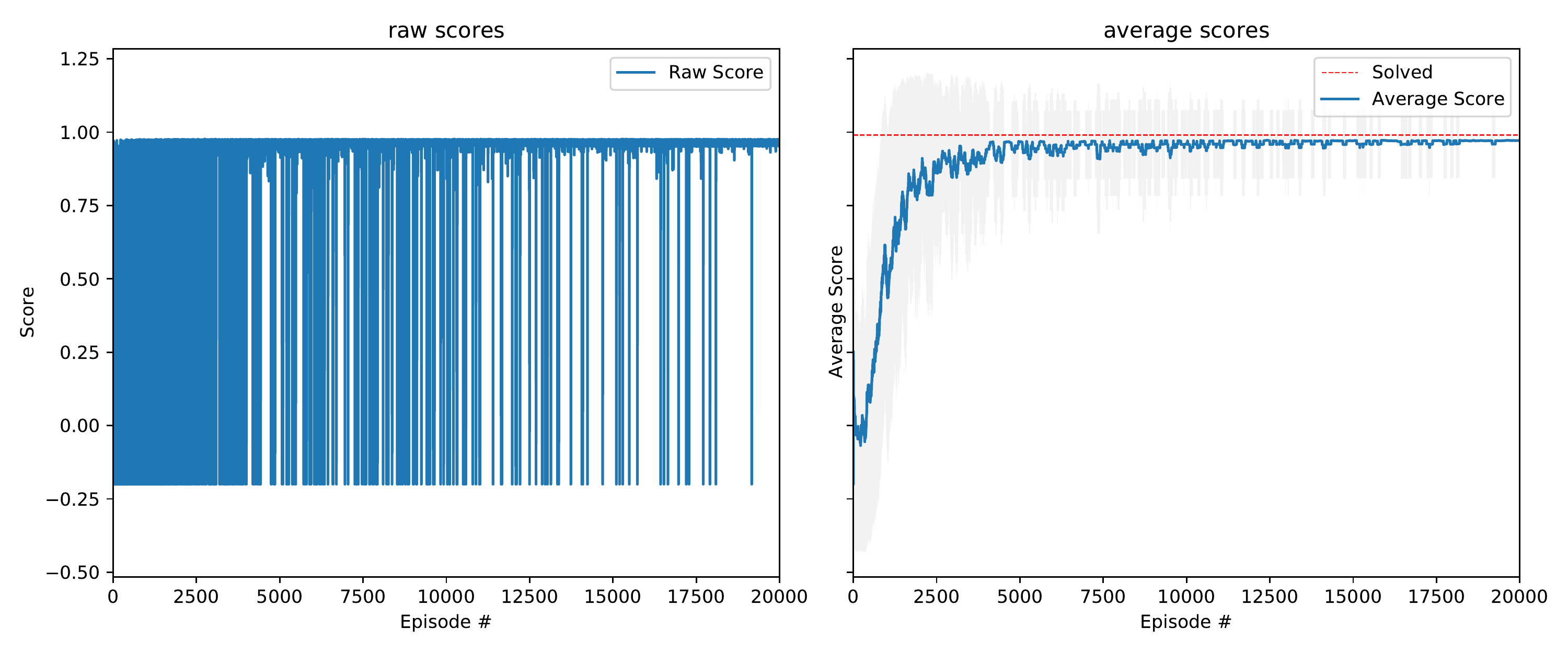}% Here is how to import EPS art
            }
        \caption{\label{fig:noisy_two_qubits_ppo_noise0001_fidelity_threshold_099} {\bfseries PPO for noisy two-qubit system with single-qubit error rate $0.001$ and fidelity threshold $0.99$.}}
     \end{subfigure}
     \hfill
     \begin{subfigure}[t]{1\textwidth}
     \centering
         \scalebox{0.4}{
            \includegraphics{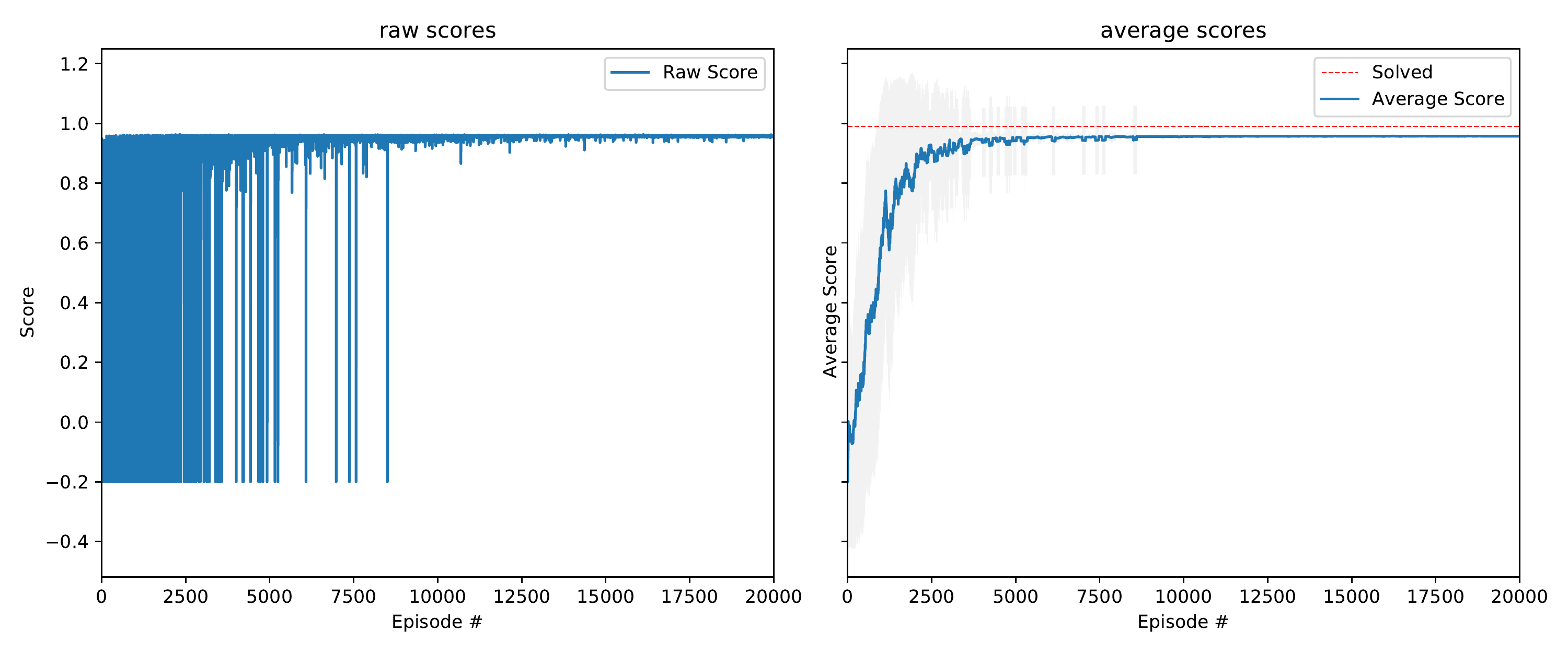}% Here is how to import EPS art
            }
        \caption{\label{fig:noisy_two_qubits_ppo_noise0005_fidelity_threshold_095} {\bfseries PPO for noisy two-qubit system with single-qubit error rate $0.005$ and fidelity threshold $0.95$.}}
     \end{subfigure}
     \hfill
    \caption{{\bfseries Deep Reinforcement for Noisy Two-Qubit system.} Synthesis of the Bell state.}
    \label{fig:noisey_two_qubit}
\end{figure}

In the previous section, we observe that the RL training based on PPO algorithm converges faster. In the noisy scenario, we only use the PPO and not the vanilla A2C since that the noisy environment is considered harder than the noise-free one. 
Here we study the case of applying DRL agent to synthesize the  $2$-qubit Bell state under noisy environment.
%
% \YC{setting of the experiments}
% \YC{need to specify the maximum steps 20??}
The first case we consider is with single-qubit error rate $ = 0.001$ and the fidelity threshold $ = 0.95$. Similar to the previous noise-free $2$-qubit experiments, the agent gets a negative reward $-0.01$ at each step to encourage the shortest path. The maximum steps an agent can try in an episode is still $20$. If the agent can reach fidelity beyond the threshold $0.95$, then the agent will receive a positive reward (fidelity$ - 0.01$). Otherwise it will only receive reward $ = -0.01$ when the episode ends. The result is shown in \figureautorefname{\ref{fig:noisy_two_qubits_ppo_noise0001_fidelity_threshold_095}}. Compared to the setting with the same single-qubit error rate and fidelity threshold $= 0.99$ (shown in \figureautorefname{\ref{fig:noisy_two_qubits_ppo_noise0001_fidelity_threshold_099}}), we observe that the one with fidelity threshold $=0.95$ performs better, with much more stable score (smaller standard deviation).
%
% \YC{potential reason?}

Here we need to point out that the fidelity threshold is to define whether the agent reaches a minimum goal. The agent is still trained to maximize the overall return, and the final fidelity which the agent can achieve is not limited to this threshold.
A potential explanation is that, under the setting of fidelity threshold $=0.95$, the agent would receive more guidance in the training phase. If the threshold is high, say $0.99$, then the agent will stop after the maximum attempts and get no information about the fidelity in many of the training episodes. On the other hand, if the fidelity threshold is lower, the agent would receive positive reward in more training episode, which will in turn help the agent to adjust its model parameters.

Finally we compare the performance between single-qubit error rate $ = 0.001$ and $0.005$, both with the fidelity threshold $=0.95$. We observe that both cases (shown in \figureautorefname{\ref{fig:noisy_two_qubits_ppo_noise0001_fidelity_threshold_095}} and \figureautorefname{\ref{fig:noisy_two_qubits_ppo_noise0005_fidelity_threshold_095}}) converge quickly. However, the final converged fidelity in the case with higher error rate is a bit lower.

\section{\label{sec:Discussion}Discussion}
\subsection{Relevant Works}
%
% \textbf{Cite recent quantum architecture search and quantum control results.}
Deep reinforcement learning techniques have been applied in the investigation of quantum computing technologies. There are three main categories: quantum error correction, quantum optimal control and quantum architecture search.
%
% \textbf{Recent QAS}
Early works on the quantum architecture search based on the heuristic search methods \cite{williams1998automated}. Recent works focus on the application of machine learning techniques \cite{pirhooshyaran2020quantum, du2020quantum, zhang2020differentiable, wu2020optimizing, zhang2021neural}. Our methods differ from these. For example, in the work \cite{wu2020optimizing,du2020quantum}, the authors proposed frameworks to optimize the existing quantum circuit, reducing the number of quantum gates. In our method, there is no existing quantum circuit to be optimized. The quantum circuit is to be generated from scratch. Our method is also different from \cite{zhang2020differentiable} as we do not directly sample from a distribution of quantum circuits. Our method is also different from \cite{pirhooshyaran2020quantum} as we are not to generate a batch of circuits in each time step nor using random search, instead, we would let the circuit grow incrementally. Recent work on optimizing parameterized quantum structures \citep{ostaszewski2021structure} indicates potential direction of extending AI/ML method to a more general setting (e.g. optimizing the quantum circuit architecture and its parameters simultaneously). %\textcolor{red}{Leo: what does the last sentence mean?} \YC{I mean, to automatically generate the circuit architecture AND their circuit parameters!}
\subsection{More Complex Problems}
One may wonder how efficient this approach can be extended to large quantum circuits. In general, this should be very difficult and the complexity scales exponentially in $N$, the number of qubits. 
%However, in principle, there are finite sets of one and two-qubit gates that are universal in an approximate sense \citep{kitaev1997quantum,  harrow2002efficient}. \textcolor{red}{<-- Leo: I don't understand this sentence} \textcolor{green}{En-Jui: this sentence just means that we only need finite sets of gates to approximate any quantum state within arbitrary error.} 
However, given a universal set of one and two-qubit quantum gates, in principle to approximate any quantum state (up to an error tolerance) it only requires a finite number of gates.
So our approach is still valid for arbitrary large qubit size but computationally hard. There is no free lunch we can get. But our method is still useful to construct the general density matrix. 
% \textcolor{red}{Leo: I don't understand if we claim finding quantum circuits is hard, why finding density matrices can be ``easy''?} 
% \textcolor{green}{En-Jui: I think what I want to say is that in general case, finding a circuit to prepare state is computationally hard. But for the usual purpose like creating the common state, our scheme still works easily. Theoretically speaking, the quesiton is hard from complexity point of view, but for the practical use, it looks not so difficult at least in our cases.} 
It is interesting to see the difficulty in the training related to the complexity of the target density matrix. There are various ways to define the complexity of the density matrices \citep{plenio2005logarithmic, vidal2002computable},
the connection to which we leave as future work.
\subsection{Noisy Environments}
In this work we investigate the potential of applying deep reinforcement learning in the quantum gate search under simple noisy configurations. Our proposed software toolkit and framework is possible to be extended into other more complex noise models. Therefore, the results of this work is a good choice of testbed for a variety of future studies concerning different noise or error schemes. For example, recent works suggest that ML models can be used to learn the quantum circuit architecture under the noise effects \cite{cincio2021machine}. We expect our framework can be incorporated with such techniques. In addition, real quantum computers have different hardware topologies and these indeed have influences on the circuit design, we leave these topology-aware quantum architecture search as future work.

% \textbf{constrained/topology-aware search}
%
\subsection{Real Quantum Computers}

It is interesting to ask whether one can use real quantum computers to realize our algorithms. Our platform is based on Qiskit. Therefore, one can easily connect our module by connecting the real IBM quantum computer. Thousand of training episodes are required in our experiment, however, real quantum computers do not have so much resource to do. So it is interesting to investigate this problem when quantum computing resources are more accessible. We leave it as future work.
\subsection{Other Quantum States}
In this work, we consider the cases of two-qubit Bell state and three-qubit GHZ state for demonstration. However, the framework of the testing environment and RL agents are rather general. It is possible to investigate the quantum architecture search problem with other target quantum states and different noise configurations. In addition, it is also very convenient to test the performance of different reinforcement learning algorithms on quantum architecture search via the standard OpenAI Gym interface.
\subsection{Extension of the Environment}
In this work, we pre-defined a set of gates or operations which can be used to generate a desired quantum state. This is not a limitation of our framework. The testing environment itself can be extended or modified to fit the quantum computing devices that are of interest. For example, it is interesting to build customized training environments with available operations from a specific quantum hardware.
\subsection{Circuit Optimization}
% \YC{check the bib form here}
One relevant question is that once we give a circuit $C$ to produce a particular quantum state. Can one optimize this circuit getting a new circuit $C'$ by reducing its depth and circuit complexity? The answer is yes, recently, there is a paper using reinforcement learning for given a circuit representation \cite{fosel2021quantum} and optimize the circuit depth. So \cite{fosel2021quantum} can be viewed as the next step or the useful tool to optimize our circuits. However, we are building quantum circuit from scratch. So our goal is different from theirs. One can easily see from the complexity point of view, solving our task efficiently does not imply solving their task efficiently and vice versa. One can indeed combine our work and their work to form a pipeline to solve the following: given a target state $\psi$ and then try to find an efficient circuit $C$ such that using $C$ to create the target state. There is another related paper \cite{ostaszewski2021reinforcement} using reinforcement learning and variational quantum circuit to find the ground state.
\section{\label{sec:Conclusion}Conclusion}
In this work, we demonstrate the application of deep reinforcement learning (DRL) to automatically generate the quantum gates sequence from the density matrix only. Our results suggest that with the currently available deep reinforcement learning algorithms, it is possible to discover the near-optimal quantum gate sequence with very limited physics knowledge encoded into the RL agents. We also present the customized OpenAI Gym environment for the experiments, which is a valuable tool for exploring other related quantum computing problems.

\begin{acknowledgments}
This work is supported by the U.S.\ Department of Energy, Office of Science, Office of High Energy Physics program under Award Number DE-SC-0012704 and the Brookhaven National Laboratory LDRD \#20-024.
\end{acknowledgments}
% \paragraph*{S.Y.-C.C. and E.-J.K. contribute equally to this work.}

\appendix

\section{RL Algorithms}
Here we provide the details of the RL algorithms used in this work. 
\begin{center}
\scalebox{0.9}{
\begin{minipage}{\linewidth}

\begin{algorithm}[H]
\begin{algorithmic}

\State Define the number of total episode $M$
\State Define the maximum steps in a single episode $S$

\For{episode $=1,2,\ldots,M$}
    \State Reset the testing environment and initialise state $s_1$
    \State Initialise trajectory buffer $\mathcal{T}$
    \State Initialise the counter $t$
    \State Initialise episode reward $R_E = 0$
    \For{step $=1,2,\ldots,S$}
        \State Select the action $a_t$ from the policy $\pi\left(a_{t} \mid s_{t} ; \theta_{\pi}\right)$
        
        \State Execute action $a_t$ in emulator and observe reward $r_t$ and next state $s_{t+1}$
        
        \State Record the transition $\left(s_t,a_t,r_t,s_{t+1}\right)$ in $\mathcal{T}$
        
        \State Episode reward $R_E \leftarrow R_E + 1$
        
        \If{reaching terminal state \OR reaching maximum steps $M$}
            % \State $R=\left\{\begin{array}{ll}0 & \text { for terminal } s_{t} \\ V\left(s_{t}, \theta_{v}^{\prime}\right) & \text { for non-terminal } s_{t} \end{array}\right.$
            % \For{$i \in\left\{t-1, \ldots, t_{0}\right\}$}
            %     \State $R \leftarrow r_{i}+\gamma R$
            %     % \State 
            % \EndFor
            \State Calculate the value targets $R_{t}$ for each state $s_t$ in the trajectory buffer $\mathcal{T}$
            \State Calculate the values $V(s_{t}, \theta_{v})$ of each state $s_t$ from the model $V\left(s_{t}, \theta_{v}\right)$
            \State Calculate the value loss $L_{\text{value}} = \mathbb{E}_{t} \norm{V(s_{t}, \theta_{v}) - R_{t}}^{2} $
            \State Calculate the entropy term $H = \sum_{t} H_t = \sum_{t} \left[-\sum_{j} \pi\left(a_{j} \mid s_{t} ; \theta_{\pi}\right) \log(\pi\left(a_{j} \mid s_{t} ; \theta_{\pi}\right)) \right]$

            \State Calculate the advantage $A_t = R_{t} - V(s_{t}, \theta_{v})$
            \State Calculate the policy loss $L_{\text{policy}} = \mathbb{E}_{t} \left[- \log \pi\left(a_{t} \mid s_{t} ; \theta_{\pi}\right) A_{t} \right]$
            \State Total loss $L = L_{\text{value}} + L_{\text{policy}} - 0.001 \times H$
            \State Update the agent policy parameters $\theta_{\pi}$ and $\theta_{v}$ with gradient descent on the loss $L$
        \EndIf

    \EndFor

\EndFor
\end{algorithmic}
\caption{Advantage Actor-Critic (A2C) for quantum architecture search}
\label{a2c_quantum_gates}
\end{algorithm}
\end{minipage}
}
\end{center}
\begin{center}
\scalebox{0.9}{
\begin{minipage}{\linewidth}

\begin{algorithm}[H]
\begin{algorithmic}
\State Define the number of total episode $M$
\State Define the maximum steps in a single episode $S$
\State Define the update timestep $U$
\State Define the update epoch number $K$
\State Define the epsilon clip $C$
\State Initialise trajectory buffer $\mathcal{T}$
\State Initialise timestep counter $t$
\State Initialize two sets of model parameters $\theta$ and $\theta_{\text{old}}$
\For{episode $=1,2,\ldots,M$} 
    \State Reset the testing environment and initialise state $s_1$
    \For{step $=1,2,\ldots,S$}
        \State Update the timestep $t = t + 1$
        
        \State Select the action $a_t$ from the policy $\pi\left(a_{t} \mid s_{t} ; \theta_{\text{old}}\right)$
        
        \State Execute action $a_t$ in emulator and observe reward $r_t$ and next state $s_{t+1}$
        
        \State Record the transition  $\left(s_t,a_t,\log \pi\left(a_{t} \mid s_{t} ; \theta_{\text{old}}\right) ,r_t\right)$ in $\mathcal{T}$
        
        \If{$t = U$}
            \State Calculate the discounted rewards $R_{t}$ for each state $s_t$ in the trajectory buffer $\mathcal{T}$
            
            \For{$k = 1,2,\ldots,K$}
                \State Calculate the log probability $\log \pi\left(a_{t} \mid s_{t} ; \theta \right)$, state values $V(s_{t},\theta)$ and entropy $H_{t}$.
                
                \State Calculate the ratio $ q_{t} = \exp{(\log \pi\left(a_{t} \mid s_{t} ; \theta \right) - \log \pi\left(a_{t} \mid s_{t} ; \theta_{\text{old}} \right))}$
                
                \State Calculate the advantage $A_{t} = R_{t} - V(s_{t},\theta) $
                
                \State Calculate the $surr_1 = q_{t} \times A_{t}$ 
                
                \State Calculate the $surr_2 = clip(q_{t}, 1 - C, 1 + C) \times A_{t}$
                
                \State Calculate the loss $L = \mathbb{E}_{t}[-min(surr_1,surr_2) + 0.5 \norm{V(s_t,\theta) - R_{t}}^{2} - 0.01 H_{t}]$
                
                \State Update the agent policy parameters $\theta$ with gradient descent on the loss $L$
                
            \EndFor
            
            \State Update the $\theta_{\text{old}}$ to $\theta$
            \State Reset the trajectory buffer $\mathcal{T}$
            \State Reset the timestep counter $t = 0$
        \EndIf

    \EndFor
    
\EndFor
\end{algorithmic}
\caption{PPO for quantum architecture search}
\label{ppo_quantum_gates}
\end{algorithm}
\end{minipage}
}
\end{center}
\section{Code samples}
Consider the case of noise-free two-qubit system, the OpenAI Gym environment setting is as follows,
\begin{minted}{python}
import gym
import gym_twoqubit
target = np.asarray([0.70710678+0.j,0.        +0.j,0.        +0.j, 0.70710678+0.j])
env = gym.make('BasicTwoQubit-v0', target = target)
\end{minted}
where we import relevant packages and set the target of the quantum state that we want the RL agent to learn. The target is used to initialize the gym environment.
Consider the case of noisy two-qubit system, the OpenAI Gym environment setting is as follows,
\begin{minted}{python}
import gym
import gym_twoqubit
from qiskit.providers.aer.noise import NoiseModel
from qiskit.providers.aer.noise.errors import pauli_error, depolarizing_error

def get_noise(p_meas,p_gate):
	error_meas = pauli_error([('X',p_meas), ('I', 1 - p_meas)])
	error_gate1 = depolarizing_error(p_gate, 1)
	error_gate2 = error_gate1.tensor(error_gate1)

	noise_model = NoiseModel()
	# measurement error is applied to measurements
	noise_model.add_all_qubit_quantum_error(error_meas, "measure")
	# single qubit gate error is applied to x gates
	noise_model.add_all_qubit_quantum_error(error_gate1, ["x"]) 
	# two qubit gate error is applied to cx gates
	noise_model.add_all_qubit_quantum_error(error_gate2, ["cx"]) 
		
	return noise_model

def generate_backend_noise_info(backend):
	device_backend = backend
	coupling_map = device_backend.configuration().coupling_map
	noise_model = NoiseModel.from_backend(device_backend)
	basis_gates = noise_model.basis_gates

	backend_noise_info = {
	"noise_model": noise_model,
	"coupling_map": coupling_map,
	"basis_gates": basis_gates,
	}

	return backend_noise_info

noise_model = get_noise(0.001,0.001)
backend_noise_info = backend_noise_info = {
"noise_model": noise_model,
"coupling_map": None,
"basis_gates": None,
}
target = np.asarray([0.70710678+0.j,0.        +0.j,0.        +0.j, 0.70710678+0.j])
env = gym.make('NoisyTwoQubit-v0',
target = target, 
backend_noise_info = backend_noise_info, 
verbose = True, 
fidelity_threshold = 0.99)
\end{minted}
where we import relevant packages and set the target of the quantum state that we want the RL agent to learn. In addition, we use functions from Qiskit package to define the noise model and the quantum simulation backend. The target and backend setting are then used to initialize the gym environment. 
We adopt the code for generating noise model from IBM qiskit textbook \cite{qiskittext}.
\bibliographystyle{ieeetr}
\bibliography{apssamp,bib/classical_rl,bib/qas_related,bib/nas_related,bib/tool,bib/nisq,bib/qecc_related,bib/quantum_control_related,bib/quantum_optimization_related,bib/ml_quantum_error_mitigation,bib/qml_examples,bib/qc,bib/vqc}% Produces the bibliography via BibTeX.

\end{document}